# Operationalizing AI: Empirical Evidence on MLOps Practices, User Satisfaction, and Organizational Context

Stefan Pasch[1]

**Abstract**

Organizational efforts to utilize and operationalize artificial intelligence (AI) are often accompanied by substantial challenges, including scalability, maintenance, and coordination across teams. In response, the concept of *Machine Learning Operations (MLOps)* has emerged as a set of best practices that integrate software engineering principles with the unique demands of managing the ML lifecycle. Yet, empirical evidence on whether and how these practices support users in developing and operationalizing AI applications remains limited. To address this gap, this study analyzes over 8,000 user reviews of AI development platforms from G2.com. Using zero-shot classification, we measure review sentiment toward nine established MLOps practices, including continuous integration and delivery (CI/CD), workflow orchestration, reproducibility, versioning, collaboration, and monitoring. Seven of the nine practices show a significant positive relationship with user satisfaction, suggesting that effective MLOps implementation contributes tangible value to AI development. However, organizational context also matters: reviewers from small firms discuss certain MLOps practices less frequently, suggesting that organizational context influences the prevalence and salience of MLOps, though firm size does not moderate the MLOps-satisfaction link. This indicates that once applied, MLOps practices are perceived as universally beneficial across organizational settings.

## 1. Introduction

The past decade has seen a rapid rise in the adoption and application of artificial intelligence (AI) across industries. From generative models to predictive analytics, AI technologies increasingly shape how organizations create value, optimize processes, and engage with customers. As a result, AI has become not only a topic of technological innovation but also a strategic priority for organizations seeking to integrate it into products, services, and decision-making processes. However, many organizational efforts to implement and scale AI systems face substantial challenges, with a large share of projects failing to move beyond pilot stages and to generate sustained business value (Ryseff et al., 2024; Westenberger et al., 2022). In particular, as these systems scale, organizations encounter increasing complexity in managing the development, deployment, and maintenance of ML models (Shivashankar et al., 2025).

In response, *Machine Learning Operations* (MLOps) has emerged as a set of best practices that integrate software engineering principles with the unique requirements of ML system management – spanning automation, reproducibility, monitoring, collaboration, and continuous delivery (Testi et al., 2022). MLOps is increasingly seen as essential for ensuring the reliability, scalability, and governance of AI systems (Eken et al., 2024).

---

[1] Division of Social Science & AI, Hankuk University of Foreign Studies: stefan.pasch@outlook.com

Despite its rapid diffusion in both research and practice, our empirical understanding of MLOps remains limited. Much of the existing literature is conceptual, architectural, or tool-oriented, focusing on defining the MLOps lifecycle, outlining technical frameworks, or surveying available tools and platforms (Kreuzberger et al., 2023; Zarour et al., 2025). Empirical evidence on how MLOps practices influence user or organizational outcomes is still scarce. In particular, little is known about how MLOps practices support those who develop and operationalize AI systems in their daily work.

To address this gap, we analyze more than 8,000 user-generated reviews of AI and ML development platforms from G2.com, a leading business software review platform. These platforms, which include data science environments, ML lifecycle tools, and MLOps solutions, are central to how organizations build and deploy AI. Using zero-shot text classification with a large language model (Llama 3.3 70B), we measure the sentiment on nine established MLOps practices, such as CI/CD automation, workflow orchestration, reproducibility, versioning, collaboration, and continuous monitoring, based on the free-text content of user reviews. In turn, we evaluate how MLOps practices relate to overall user satisfaction, an established indicator of perceived system quality and success (DeLone & McLean, 2003).

Across nine established MLOps practices, we find consistent evidence that these practices are positively associated with user satisfaction, with all but two – CI/CD automation and versioning – showing statistically significant effects. This supports the view that MLOps practices meaningfully contribute to the development and operationalization of AI, rather than representing a purely technical ideal or industry trend. These findings underscore the importance of MLOps as a socio-technical capability that enhances users' experience of building and deploying machine learning systems.

However, MLOps practices may not matter equally for all users engaged in the development and deployment of AI. The organizational context in which these practices are applied is likely to influence their salience and perceived value. Because MLOps practices are designed to reduce the complexity of building, deploying, and maintaining ML systems, their benefits may grow with the scale and coordination demands of the organization. To capture this contextual dimension, we focus on firm size – a well-established proxy for organizational complexity, resource availability, and coordination needs in information systems research (Giunta & Trivieri, 2007; Na et al., 2023). We analyze how firm size of reviewers shape both the extent to which MLOps practices are discussed and the strength of their relationship with user satisfaction.

While our results show that reviewers from smaller firms discuss certain MLOps practices less frequently than those working for large enterprises, the overall satisfaction benefits of these practices appear largely consistent across firm sizes. This suggests that firm size influences the *visibility* and *adoption* of MLOps practices more than their *intrinsic value* once in use – highlighting that the benefits of MLOps extend broadly across organizational contexts once users experience them.

Overall, this study makes three key contributions. First, it introduces a scalable, language-model–based approach to measuring MLOps practices from unstructured text, enabling systematic empirical study of socio-technical processes that were previously difficult to observe. Second, it provides large-scale evidence that MLOps practices are positively associated with user satisfaction, demonstrating their practical relevance for the development and operation of AI systems. Third, it shows how organizational context – captured through firm size – shapes engagement with MLOps practices: while smaller firms refer to them less frequently, their satisfaction benefits remain consistent across organizational scales.

## 2. Theoretical Background

### 2.1. MLOps Practices and User Satisfaction

The development and deployment of machine learning (ML) systems has undergone a fundamental transformation in recent years. In the early phases of industrial ML adoption, workflows were largely experimental and ad hoc, often built around isolated Jupyter notebooks or scripts, with minimal integration into broader engineering or production infrastructures (Sculley et al., 2015; Polyzotis et al., 2018). While these lightweight environments supported rapid prototyping, they also introduced significant limitations in terms of scalability, reproducibility, and operational reliability – particularly when models were transitioned into production or collaborative settings (Zhao et al., 2024).

To address these limitations, the field has increasingly embraced Machine Learning Operations (MLOps) as a set of best practices, technologies, and organizational principles aimed at supporting the full ML lifecycle. MLOps focuses on automating and managing the processes of model development, deployment, monitoring, and maintenance in a scalable and repeatable manner (Zarour et al., 2025; Kreuzberger et al., 2023). Key practices include continuous integration and delivery (CI/CD) of ML models, reproducibility and version control for data and experiments, collaborative workflows, experiment metadata tracking, continuous training, feedback loops from deployment back to training, and ongoing monitoring of model behavior. These practices are often embedded into end-to-end platforms that support not just model development, but also the complex coordination required for real-world deployment and maintenance.

Despite the growing importance of MLOps, there is limited empirical research on how users perceive these practices in practice, and how such perceptions influence their satisfaction with ML platforms. In this study, we aim to fill this gap by examining the relationship between sentiment toward MLOps practices and user satisfaction as expressed in real-world product reviews.

User satisfaction has long been recognized as a central outcome in Information Systems (IS), Human-Computer Interaction (HCI), and technology management research. It serves as a key proxy for system success and perceived value, capturing both the user's cognitive evaluation of a platform's effectiveness and the affective responses it elicits (DeLone & McLean, 2003; Bhattacherjee, 2001). In both organizational and individual use contexts, satisfaction reflects whether a system meets users' expectations, supports their work goals, and facilitates meaningful outcomes. We argue that the ability to effectively integrate and execute MLOps best practices is a central driver of user satisfaction with machine learning and data science platforms.

There are various potential channels through which MLOps practices could affect user satisfaction. First, MLOps practices improve deployment speed and reliability. Techniques like CI/CD automation, automated testing, and model performance monitoring reduce friction in the release process and minimize errors in transitioning from development to production. Faster, more reliable deployment allows teams to respond quickly to changing requirements or model degradation, which is particularly critical in dynamic or data-intensive domains. Research from DevOps and software engineering shows that higher deployment velocity and stability are associated with increased user and developer satisfaction (Forsgren et al., 2019). Similarly, IS research links timely updates and low failure rates to higher perceived system quality (Li & Zhu, 2022).

Second, MLOps supports operational efficiency by automating workflows and fostering collaboration. Practices such as pipeline orchestration, metadata tracking, and shared dashboards reduce duplicated effort, clarify task responsibilities, and make team coordination more seamless. This aligns with longstanding IS and HCI research emphasizing that tools aligned with user workflows and team structures are more likely to be perceived as usable and effective (Ren et al., 2023). When platforms enable cross-functional teams – including data scientists, ML engineers, and business users – to work together with minimal friction, satisfaction tends to increase (Espinosa et al., 2007; Ahmad et al., 2023).

Third, MLOps helps users manage risk and reduce uncertainty. Reproducibility practices, such as experiment tracking, data versioning, and artifact management, help ensure consistency across experiments and environments. Continuous monitoring enables users to catch issues like data drift, bias, or performance degradation after deployment. These capabilities reduce the perceived risk of unexpected model failures – a key concern in high-stakes domains. Risk reduction is known to improve perceived control and trust in digital systems (Pavlou, 2003; McKnight et al., 2002), and this logic extends to ML systems where consequences of failure can be substantial.

Fourth, MLOps practices can reduce cognitive load and increase task clarity. Structured pipelines, automation of repetitive steps, and clear interfaces (e.g., workflow visualizations) help users focus on core analytical or decision-making tasks instead of dealing with low-level infrastructure. In HCI, reducing cognitive complexity is known to enhance user satisfaction (Gudigantala et al., 2011; Schmidhuber et al., 2021), especially for users operating under time pressure or across multiple tools. MLOps platforms that provide clarity in the model lifecycle – from data ingestion to retraining – are thus more likely to be viewed positively.

Fifth, MLOps enhances transparency and explainability. Features such as logging, metadata tracking, and feedback mechanisms help users trace model lineage, audit decisions, and understand performance changes over time. Transparency is a known contributor to trust, control, and satisfaction in both IS and AI system research (Rai, 2020; Shneiderman, 2020). When users can understand why a model behaves a certain way, they are more likely to feel empowered and capable.

Taken together, these five channels offer a robust theoretical basis for expecting a positive relationship between perceived MLOps practices and user satisfaction. In our study, we measure sentiment toward these practices using a large-scale analysis of platform reviews and examine their association with satisfaction ratings. We propose the following hypothesis:

***H1: Sentiment on MLOps practices is positively associated with user satisfaction.***

## 2.2. Salience of MLOps Practices and Firm Size

While MLOps best practices are expected to enhance the development and deployment of AI and machine learning models – and thereby to increase user satisfaction – their application and perceived importance can vary with organizational context. Because MLOps practices are designed to reduce the complexity of building, deploying, and maintaining ML systems, their benefits are likely to grow with the scale and coordination demands of the organization in which they are applied. Among the many structural attributes of firms, size stands out as a well-established proxy for the scale and complexity of digital operations. In both management and information systems research, firm size is frequently used to capture differences in resources, coordination demands, and governance requirements that shape how technology is implemented and experienced (Giunta & Trivieri, 2007; Na et al., 2023). Building on this tradition, we examine firm size as a key contextual factor to understand how MLOps practices are discussed and valued.

As firms grow in size, their machine learning operations typically become more complex: the number of deployed models increases, data pipelines multiply, and development involves a wider set of engineers, data scientists, and business stakeholders. This escalation in scope creates higher demands for coordination, reliability, and compliance – precisely the challenges that MLOps practices are designed to address. Building on this observation, several theoretical perspectives suggest why larger firms may be more likely to notice and value these practices when evaluating machine learning platforms.

First, information processing theory argues that organizations must balance the volume of information they generate with their capacity to process it (Galbraith, 1973). As firms expand, the sheer number of interdependent ML tasks raises information load and uncertainty. To cope, larger firms benefit from formalized processes that reduce complexity and risk. Complementary frameworks such as the Technology–Organization–Environment model and Diffusion of Innovation theory likewise find firm size to be a robust predictor of structured technology adoption, as larger firms have more formal processes and stronger incentives to implement technologies that can handle scale (Oliveira & Martins, 2011; AL-Shboul, 2019). Similarly, empirical research on enterprise systems and workflow automation suggests that larger firms are earlier and heavier adopters of practices designed to coordinate complex, cross-functional work (Zhu et al., 2006). These findings suggest that MLOps, which aims to manage complex machine learning pipelines, will be especially central and visible in large organizations.

In addition, task–technology fit theory (Goodhue & Thompson, 1995) highlights that the value of a technology depends on how well its functionality matches the requirements of the tasks it supports. As machine learning tasks become more demanding – encompassing continuous model updates, integration with multiple services, and high reliability expectations – MLOps practices such as CI/CD automation, reproducibility, and continuous monitoring become indispensable. Prior IS research shows that complex, interdependent tasks strengthen the fit between advanced IT systems and user needs, increasing user recognition and discussion of those systems' capabilities (Strong et al., 2006; Zigurs & Buckland, 1998). Larger firms, which typically operate with more complex ML tasks and higher stakes in production reliability, therefore provide a context in which MLOps practices are not just beneficial but integral to daily operations.

Finally, the resource-based view (Barney, 1991) highlights that larger firms generally possess more financial, technical, and human capital. These resources not only enable investment in robust infrastructure and formal governance mechanisms, but also support training programs, dedicated teams, and routinized processes for adopting new practices. Empirical evidence shows that larger firms not only provide more extensive and systematically evaluated employee training (Asadullah et al., 2015) but also invest more heavily in research and development (Shefer & Frenkel, 2005), enabling them to build stronger technological capabilities and to implement IT best practices and other organizational innovations.

Taken together, these perspectives converge on the expectation that reviewers from larger firms are more likely to encounter, depend on, and explicitly remark upon MLOps practices when evaluating ML platforms. Therefore, we hypothesize:

***H2: Reviewers from larger firms are more likely to discuss MLOps practices in their evaluations of ML platforms.***

### 2.3. Firm Size as a Moderator of the MLOps–Satisfaction Link

While H2 focuses on how firm size influences the *salience* of MLOps practices – how often and explicitly users discuss them – firm size may also shape the *strength* of the link between sentiment toward those practices and user satisfaction. In other words, firm size might not only affect whether MLOps is noticed but also how useful and beneficial it is perceived to be once present.

Many of the arguments developed for H2 can also inform this moderating perspective. Information processing theory (Galbraith, 1973) and task–technology fit (Goodhue & Thompson, 1995) both emphasize that when organizational tasks are complex and interdependent, technologies that formalize and automate workflows create greater marginal value. As firms grow, their machine learning activities typically involve larger and more heterogeneous datasets, more frequent and parallel model deployments, and tighter interdependencies among engineering, data science, and product teams. Under these conditions, MLOps practices are not only more frequently applied and more salient, as argued in H2, but are also perceived as more beneficial for achieving dependable and scalable AI operations. From a task–technology fit perspective, the alignment between what MLOps practices offer – such as CI/CD automation, reproducibility, and continuous monitoring – and the complexity of the tasks being performed strengthens with firm size. Consequently, positive sentiment toward these practices is likely to translate into especially strong satisfaction when the organizational setting makes them indispensable to everyday machine learning work.

While larger firms may find MLOps practices more salient and more frequently applied, an alternative perspective suggests that once these practices are adopted and noticed, their satisfaction benefits are largely universal. From a resource-based view (Barney, 1991), firm size mainly determines an organization's ability to *adopt* and maintain advanced practices, not how beneficial those practices are after adoption. In this sense, firm size shapes exposure rather than the intrinsic value of MLOps once experienced.

Moreover, many of the benefits of MLOps are not contingent on large-scale or highly complex operations. Practices such as reproducibility, versioning, and metadata tracking prevent accidental errors, ensure traceability, and facilitate debugging – capabilities that can be equally valuable for a single data scientist working on a small project as for a large enterprise team. Continuous monitoring and automated feedback loops help detect model drift or data quality issues regardless of whether a model serves millions of users or a narrow internal application. In line with IS success and technology-acceptance research, core qualities such as reliability, ease of use, and transparency typically drive satisfaction across organizational contexts once users perceive them (DeLone & McLean, 2003; Davis, 1989).

Taken together, the theoretical arguments do not point to a single conclusion about whether firm size strengthens or leaves unchanged the MLOps–satisfaction relationship. This ambiguity gives rise to two competing hypotheses:

*H3a: The positive association between sentiment toward MLOps practices and user satisfaction is stronger for reviewers from larger firms than for those from smaller firms.*

*H3b: The positive association between sentiment toward MLOps practices and user satisfaction does not differ significantly by firm size.*

## 3. Methodology

### 3.1. Data

We base our analysis on user-generated reviews from G2.com, a leading platform for evaluating business-to-business (B2B) software. Unlike consumer-oriented review sites such as Amazon or Trustpilot, G2 specializes in enterprise technologies, making it particularly suitable for studying tools used in professional AI and machine learning development (G2, 2024; Kevans, 2023). Recent research has also demonstrated the value of G2 data for examining technology use and perceptions, including phenomena such as human–AI interaction (Pasch & Ha, 2025).

To construct the dataset, we retrieved all reviews from two G2 product categories that directly support the creation and deployment of machine learning systems: "Data Science and Machine Learning Platforms" and "MLOps Platforms." These categories include a wide range of cloud-based and on-premise solutions for data preparation, model training, automated deployment, and lifecycle management. Focusing on these categories ensures that the products under study are designed for building and operating machine learning models rather than for purely end-user applications such as chatbots.

The resulting sample covers 229 products and 8,627 user reviews. Each review contains two separate open-text fields in which users describe (i) what they like and (ii) what they dislike about the product. We combined these two fields into a single review text, prefixing each segment with "Like:" and "Dislike:" markers to preserve the original section context.

The dataset also includes structured metadata. Each review is linked to a 1-to-5 star rating, which serves as our dependent variable measuring user satisfaction. G2 further reports the reviewer's firm size in three categories – small business (1–50 employees), mid-sized company (51–1,000 employees), and enterprise (>1,000 employees). We proxy firm size in the analyses with two dummy variables for small business and mid-sized firm, using enterprise as the reference group.

In addition, each review lists the reviewer's job title. Following Pasch and Cha (2025), we derive a control for whether the reviewer holds a technical job. Job titles were normalized by lowercasing and removing punctuation, and a review was coded technical = 1 if the title contained domain-specific keywords such as *engineer*, *developer*, *data scientist*, *ML engineer*, *MLOps*, *architect*, or *technical*; otherwise technical = 0. This variable helps separate evaluations driven by deep technical knowledge from those reflecting primarily managerial or end-user perspectives.

Finally, we include fixed effects for each product's G2 subcategory. All products in our sample fall under G2's broader Data Science and Machine Learning Platforms or MLOps Platforms umbrellas and are therefore geared toward AI development and deployment. Within these umbrellas, however, products differ in emphasis, such as Integrated Development Environments (IDE), Big Data Integration Platforms, and Generative AI Infrastructure Software. Controlling for these subcategories accounts for systematic differences in product focus and mitigates the risk that our results simply capture variation in niche functionality rather than MLOps-related perceptions.

## 3.2. Zero-Shot Classifications of MLOps Practices

**Table 1: Overview of MLOps Practices**

| MLOPs Practice | Description |
| --- | --- |
| P1 Automating CI/CD | Enables continuous integration, delivery, and deployment of ML models, providing rapid feedback on build and deployment success to improve productivity and reliability. |
| P2 Orchestrating workflows | Coordinates and automates the sequence of tasks in an ML pipeline, managing dependencies and execution order. |
| P3 Ensuring reproducibility | Ensures that ML experiments can be exactly repeated, supporting reliability and scientific traceability. |
| P4 Versioning | Tracks versions of data, code, and models to guarantee reproducibility and allow auditing or rollback. |
| P5 Ensuring collaboration and communication | Facilitates joint work on models, code, and data, and fosters cross-role communication to reduce organizational silos. |
| P6 Continuous training and evaluation | Supports periodic model retraining and systematic evaluation so that models stay accurate as data or conditions change. |
| P7 Tracking and logging ML metadata | Logs metadata (e.g., training parameters, performance metrics) to document model lineage and enable full traceability. |
| P8 Continuous monitoring | Continuously checks data, models, and infrastructure for drift, errors, or performance degradation to maintain product quality. |
| P9 Implementing feedback loops | Integrates insights from monitoring and evaluation back into data engineering and model development for iterative improvement. |

Note. Descriptions are based on Kreuzberger et al. (2023) and Zarour et al. (2025).

As our basis for MLOps practices we rely on the set of nine MLOps best practices that both Kreuzberger et al. (2023) and Zarour et al. (2025) identified in their literature reviews on MLOps, as shown in Table 1. These practices – ranging from CI/CD automation and workflow orchestration to reproducibility, versioning, collaboration, continuous training, metadata tracking, continuous monitoring, and feedback loops – capture the essential technical and organizational routines that enable production-ready machine learning.

To measure how these practices are discussed in user reviews, we employed zero-shot classification with the Llama 3.3 70B large language model (LLM). Zero-shot classification allows an LLM to assign texts to predefined categories without supervised training data. When provided with clear, structured instructions, the model uses its broad semantic knowledge to judge whether a review mentions each MLOps practice and whether the mention is positive, negative, or neutral. In essence, the model reasons about category membership on the fly, rather than relying on patterns learned from labeled examples.

Zero-shot LLMs have recently proven capable of near human-level text classification when given explicit instructions and structured label definitions (Törnberg, 2023; Chae & Davidson, 2023; Pasch & Cutura, 2024). In HCI and IS research, they are increasingly adopted for large-scale sentiment and content analysis of domain-specific constructs such as UX dimensions (e.g., Pasch et al., 2025).

This approach offers several advantages over traditional natural language processing techniques such as word counts based on dictionaries or topic modeling (Blei et al., 2003). First, direct sentiment capture: zero-shot models can explicitly determine whether a practice is mentioned positively or negatively, instead of merely inferring tone from co-occurring words. Second, multi-label capability: a single review can simultaneously discuss multiple practices with different sentiments – something standard topic models, which typically assign one topic per text, cannot handle well. Third, contextual understanding: transformer-based LLMs leverage deep context modeling to recognize nuanced, domain-specific language (e.g., technical jargon about deployment pipelines) that classical bag-of-words or LDA methods often miss. These strengths make zero-shot LLM classification a powerful and scalable way to identify and evaluate mentions of MLOps practices in large corpora of user reviews.

We chose Llama 3.3 70B as our zero-shot classification model because Llama-family models have been shown to achieve competitive accuracy in zero- and few-shot settings across a variety of benchmarks (e.g., Touvron et al., 2023). An additional advantage is that Llama is open-source, which enables reproducibility, control over model behavior, and cost savings for large-scale application.

For each of the 8,627 reviews, we provided Llama 3.3 70B with concise definitions of the nine MLOps practices (adapted from Kreuzberger et al., 2023) and prompted it to classify each practice as positive, negative, neutral, or not discussed. Because neutral classifications were exceedingly rare (below 1% of all assignments), we combined neutral with not discussed for analysis, yielding three categories per practice (positive, negative, not discussed). Each review was thus coded with nine practice-specific sentiment values, indicating for every MLOps practice whether it was discussed positively, negatively, or not at all. The exact prompt used for the zero-shot classifications is provided in Appendix A.

**Table 2. Distribution of MLOps Practice Sentiment**

| MLOps Practice | Not Discussed | Positive Sentiment | Negative Sentiment |
|---|---|---|---|
| CI/CD | 91.7% | 8.1% | 0.1% |
| Workflows | 93.3% | 5.7% | 1.0% |
| Reproducibility | 96.7% | 1.0% | 2.3% |
| Versioning | 94.7% | 2.8% | 2.4% |
| Collaboration and communication | 52.9% | 42.2% | 4.8% |
| Cont. training and evaluation | 92.9% | 5.2% | 1.9% |
| Metadata | 98.3% | 1.1% | 0.5% |
| Cont. Monitoring | 79.1% | 2.2% | 18.6% |
| Feedback Loops | 95.9% | 1.1% | 3.0% |

Table 2 summarizes the distribution of sentiment classifications across all reviews for each of the nine MLOps practices. For every practice, we report the share of reviews with positive, negative, or no discussion. Most practices are mentioned in fewer than 10% of reviews, indicating that while MLOps is a recognized concept, only a subset of users explicitly refer to these practices in their evaluations. Collaboration and continuous monitoring stand out as the most frequently discussed dimensions – reflecting their salience in everyday ML development – whereas other practices such as metadata tracking or versioning are less commonly articulated. This uneven distribution highlights that awareness and discussion of MLOps practices vary considerably across users, underscoring the relevance of examining which organizational contexts make these practices more or less visible.

To illustrate how these automated classifications manifest in practice, Appendix B provides examples of positive and negative statements for each of the nine MLOps practices. These examples show that the zero-shot labels capture meaningful, semantically distinct aspects of how users describe MLOps-related experiences.

To further assess the accuracy of the zero-shot labels, a subset of 400 randomly selected reviews was independently annotated by one of the authors using the same nine MLOps practice definitions applied in the automated procedure. The annotator has prior experience in labeling domain-specific language across research projects in HCI, organizational research, and natural language processing (NLP). We then quantified the consistency between the model's predictions and the human annotations using Cohen's Kappa (Cohen, 1960), a statistic that adjusts for agreement occurring by chance.

Kappa values ranged from 0.72 to 0.87[2] across the nine practices, which corresponds to substantial to strong agreement under the interpretation guidelines of Landis and Koch (1977). This high level of correspondence indicates that the zero-shot Llama classifications provide a reliable approximation of human judgments and can be confidently used for large-scale empirical analyses.

## 4. Results

### 4.1. MLOps Practices and User Satisfaction

To examine how perceptions of individual MLOps practices relate to user satisfaction, we regressed reviewers' star ratings (1–5 scale) on the sentiment assigned to each of the nine practices. Sentiment was coded as –1 (negative), 0 (not mentioned), or +1 (positive). The estimated coefficients can therefore be interpreted as the average change in star rating when a given practice is discussed positively, relative to when it is not mentioned; negative mentions would imply an equivalent decrease in satisfaction.

As shown in Table 3, all nine MLOps practices are positively associated with overall user satisfaction, with seven of these relationships being statistically significant. The results provide support for H1, indicating that favorable perceptions of MLOps practices correspond to higher satisfaction with ML development platforms.

Among the significant practices, collaboration and communication exhibit the strongest effect ($\beta = 0.167$), followed by reproducibility ($\beta = 0.159$), metadata tracking ($\beta = 0.133$), continuous monitoring ($\beta = 0.103$), continuous training and evaluation ($\beta = 0.101$), feedback loops ($\beta = 0.074$), and workflow orchestration ($\beta = 0.064$). These patterns suggest that practices promoting coordination, transparency, and operational reliability are particularly salient drivers of user satisfaction.

The remaining two practices – CI/CD automation ($\beta = 0.028$) and versioning ($\beta = 0.039$) also show positive but statistically insignificant coefficients, indicating that their estimated effects are comparatively weaker and less robust.

Overall, these results support H1: positive sentiment toward MLOps best practices is systematically linked to higher satisfaction with ML development platforms.

---

[2] Individual Cohen's Kappa values were: CI/CD automation (0.80), workflow orchestration (0.73), reproducibility (0.78), versioning (0.87), collaboration (0.72), continuous training (0.79), metadata tracking (0.73), monitoring (0.72), and feedback loops (0.80)

## Table 3: MLOps Practices & User Satisfaction

| | Dep. Variable: Reviewer Rating | | | | | | | | |
|---|---|---|---|---|---|---|---|---|---|
| CICD | 0.028 (0.026) | | | | | | | | |
| Workflows | | 0.064** (0.027) | | | | | | | |
| Reproducability | | | 0.159*** (0.038) | | | | | | |
| Versioning | | | | 0.039 (0.031) | | | | | |
| Collab. | | | | | 0.167*** (0.012) | | | | |
| Cont. Training | | | | | | 0.101*** (0.027) | | | |
| Metadata | | | | | | | 0.133** (0.056) | | |
| Monitoring | | | | | | | | 0.103*** (0.017) | |
| Feedback Loops | | | | | | | | | 0.074*** (0.035) |
| Controls | Yes | Yes | Yes | Yes | Yes | Yes | Yes | Yes | Yes |
| Observations | 8627 | 8627 | 8627 | 8627 | 8627 | 8627 | 8627 | 8627 | 8627 |
| R-squared | 0.08 | 0.08 | 0.08 | 0.08 | 0.09 | 0.08 | 0.08 | 0.08 | 0.10 |

Standard errors in parentheses.* p<.1, ** p<.05, ***p<.01. Controls include: Dummies for product category, company age of reviewed product, and number of employees of reviewed product, reviewers' job role (technical vs. non-technical), reviewers' company size, and year fixed effects.

### 4.2. Firm Size and Discussion of MLOps Practices

To test H2, we examined whether reviewers from firms of different sizes differ in how often they discuss MLOps practices. Each regression in Table 4 uses a binary dependent variable indicating whether a given practice is mentioned in a review (1 = mentioned, 0 = not mentioned), regardless of whether the mention is positive or negative. Firm size is represented by two dummy variables for mid-sized firms (51–1,000 employees) and small businesses (≤ 50 employees), with enterprise firms (> 1,000 employees) serving as the reference category.

As shown in Table 4, the coefficients for firm size are predominantly negative for small businesses – with seven out of nine coefficients pointing in a negative direction – indicating that reviewers from smaller organizations mention several MLOps practices less frequently than those from large enterprises. This effect is statistically significant for four practices: CI/CD automation ($\beta = –0.017$), workflow orchestration ($\beta = –0.026$), versioning ($\beta = –0.011$), and metadata tracking ($\beta = –0.008$).

When comparing mid-sized firms to enterprises, we find no systematic differences, with only two significant effects: workflow orchestration is discussed less often ($\beta = –0.017$, $p < .01$), while monitoring is mentioned slightly more frequently ($\beta = 0.021$, $p < .10$), suggesting no systematic differences between mid-sized firms and large enterprises.

Overall, these results offer partial support for H2: reviewers from small firms discuss several MLOps practices less often, consistent with a narrower scope of MLOps implementation, whereas reviewers from mid-sized firms resemble those from large enterprises in their discussion patterns.

**Table 4: Firm Size and Discussion of MLOps Practices**

| | \multicolumn{9}{c}{Dep. Variable: Discussion of MLOps Practices} | | | | | | | | |
|---|---|---|---|---|---|---|---|---|---|
| | CI/CD | Workflow | Reproduc. | Version | Collab. | Cont. Training | Metadata | Monitoring | Feedback |
| Mid-Sized | -0.008 (0.007) | -0.017*** (0.007) | 0.005 (0.006) | 0.006 (0.006) | 0.014 (0.013) | -0.006 (0.007) | -0.000 (0.003) | 0.021* (0.011) | -0.001 (0.005) |
| Small-Sized | -0.017** (0.007) | -0.026*** (0.007) | -0.001 (0.005) | -0.011* (0.006) | 0.009 (0.013) | -0.004 (0.007) | -0.008** (0.003) | 0.0069 (0.011) | -0.004 (0.005) |
| Controls | Yes | Yes | Yes | Yes | Yes | Yes | Yes | Yes | Yes |
| Obs. | 8627 | 8627 | 8627 | 8627 | 8627 | 8627 | 8627 | 8627 | 8627 |
| R2 | 0.09 | 0.03 | 0.01 | 0.04 | 0.07 | 0.04 | 0.05 | 0.03 | 0.02 |

Standard errors in parentheses.* p<.1, ** p<.05, ***p<.01. Controls include: Dummies for product category, company age of reviewed product, and number of employees of reviewed product, reviewers' job role (technical vs. non-technical), reviewers' company size, and year fixed effects.

### 4.3. Moderating Effect of Firm Size on the MLOps–Satisfaction Relationship

To test H3a and H3b, we examined whether the relationship between sentiment toward MLOps practices and user satisfaction differs by firm size. Table 5 reports the results of moderation models in which the sentiment score for each MLOps practice is interacted with dummy variables for mid-sized firms (51–1,000 employees) and small businesses (≤ 50 employees). Again, enterprise firms (> 1,000 employees) serve as the reference category.

Across all nine practices, the interaction terms are mostly small and statistically insignificant. This indicates that the positive relationship between sentiment toward MLOps practices and user satisfaction holds consistently across firm sizes. The few significant interactions – such as the negative moderation for workflow orchestration among small firms ($\beta$ = –0.149) and the positive moderation for monitoring among mid-sized firms ($\beta$ = 0.076) – do not exhibit a systematic pattern.

Overall, these results support H3b, suggesting that firm size does not systematically alter how MLOps-related perceptions translate into satisfaction. The MLOps–satisfaction relationship appears stable across organizational contexts, implying that the perceived value of MLOps practices is broadly similar in small, mid-sized, and large firms.

Table 5: Moderating Effect on the MLOps-Satisfaction Link

| MLOps Practice | CI/CD | Workflow | Reprod. | Version | Collab. | Cont. Train. | Metadata | Cont. Monitoring | Feedback Loops |
|---|---|---|---|---|---|---|---|---|---|
| | Dep. Variable: Review Rating | | | | | | | | |
| MLOps | 0.077* | 0.119*** | 0.181*** | 0.084* | 0.163*** | 0.135*** | 0.071 | 0.070*** | 0.023 |
| | (0.040) | (0.041) | (0.061) | (0.047) | (0.020) | (0.044) | (0.087) | (0.027) | (0.058) |
| MLOps x Mid-Size | -0.104* | -0.050 | 0.079 | -0.077 | -0.008 | -0.075 | 0.037 | 0.076* | 0.138 |
| | (0.040) | (0.064) | (0.091) | (0.070) | (0.023) | (0.043) | (0.126) | (0.040) | (0.084) |
| MLOps x Small-Size | -0.0618 | -0.149** | -0.149 | -0.068 | 0.023 | -0.036 | 0.205 | 0.039 | 0.022 |
| | (0.062) | (0.067) | (0.077) | (0.077) | (0.030) | (0.062) | (0.142) | (0.040) | (0.085) |
| Controls | Yes | Yes | Yes | Yes | Yes | Yes | Yes | Yes | Yes |
| Obs. | 8627 | 8627 | 8627 | 8627 | 8627 | 8627 | 8627 | 8627 | 8627 |
| R2 | 0.08 | 0.08 | 0.08 | 0.08 | 0.09 | 0.08 | 0.08 | 0.08 | 0.08 |

Standard errors in parentheses.* p<.1, ** p<.05, ***p<.01. Controls include: Dummies for product category, company age of reviewed product, and number of employees of reviewed product, reviewers' job role (technical vs. non-technical), reviewers' company size, and year fixed effects.

## 5. Discussion

### 5.1. MLOps Practices and User Satisfaction

Although MLOps has become a central concept in AI system development, empirical evidence on whether it meaningfully affects user outcomes has been limited. In line with H1, our analysis provides clear evidence that it does. Across nine established MLOps practices, all coefficients are positive and seven are statistically significant – specifically, workflow orchestration, reproducibility, collaboration, continuous training, metadata tracking, continuous monitoring, and feedback loops. Only CI/CD automation and versioning show positive but statistically non-significant coefficients. This consistent pattern provides strong support for H1, indicating that sentiment toward MLOps practices is closely linked to how users evaluate the quality and effectiveness of ML development platforms.

These findings suggest that MLOps is not merely an abstract technical framework or industry ideal but a tangible element of user experience in machine-learning development. In this context, user satisfaction can be understood not only as a subjective evaluation of interface quality but also as reflecting how well a platform supports users in the development and operationalization of AI. Taken together, the results imply that MLOps practices create recognizable value for users. They likely do so by reducing uncertainty, improving coordination, and supporting control and transparency throughout the ML lifecycle.

The link between MLOps practices and satisfaction also carries important implications. For platform providers, meaning vendors of software that support the development and operation of AI systems – such as data science, machine learning, and MLOps platforms – the results highlight the need to both implement robust MLOps capabilities and make them visible in the user experience. Transparency features, progress feedback, and collaboration tools that foreground underlying operational quality can directly enhance perceived system value. For organizations developing or deploying AI systems, the findings suggest that MLOps should be treated as a core organizational capability. Investing in employee training, workflow integration, and tool selection that enable MLOps can enhance not only the technical performance of AI systems but also the perceived usability, transparency, and trustworthiness of the platforms that support them.

## 5.2. MLOps Practices and Firm Size

Firm size represents an important contextual factor in understanding how organizations adopt and experience MLOps practices. Our analyses reveal two distinct yet complementary patterns. First, reviewers from small firms discuss a subset of MLOps practices – namely CI/CD automation, workflow orchestration, versioning, and metadata tracking – significantly less often than those from large enterprises, providing partial support for H2. In contrast, reviewers from mid-sized organizations show few systematic differences compared to enterprise reviewers. Second, firm size does not moderate the relationship between MLOps sentiment and user satisfaction, offering support for H3b. Taken together, these findings suggest that firm size affects the *salience* of MLOps practices in user discourse but not their *impact* on satisfaction once these practices are present.

The lower frequency of MLOps-related discussions among small firms likely reflects differences in organizational complexity and maturity. Smaller companies typically operate with fewer teams, shorter communication chains, and simpler ML pipelines, which reduce the need for formalized orchestration, versioning, or metadata tracking. Notably, the practices that differ most across firm sizes – CI/CD automation, workflow orchestration, versioning, and metadata tracking – are those tied to the operationalization and infrastructure layers of MLOps, which require coordinated processes and stable tooling. In contrast, core ML development practices such as reproducibility and continuous training are discussed with similar frequency across firms, suggesting that they represent more universal aspects of machine learning work. From an information-processing perspective (Galbraith, 1973), smaller firms face fewer coordination demands and thus rely less on complex process mechanisms. Likewise, within the technology–organization–environment (TOE) framework, organizational scale and resource availability shape how systematically such advanced MLOps routines are implemented and discussed.

The absence of a significant moderating effect of firm size on the MLOps–satisfaction link suggests that, once users experience these practices, their benefits are largely universal. From a resource-based view (Barney, 1991), firm size primarily shapes an organization's ability to adopt and sustain advanced practices rather than their intrinsic value once implemented. Core MLOps capabilities – such as reproducibility, traceability, and monitoring – enhance reliability and user confidence regardless of organizational scale. These functions reduce errors and increase transparency, which are desirable whether models are developed by a single data scientist or a large enterprise team. In line with IS success and technology-acceptance research (DeLone & McLean, 2003; Davis, 1989), the drivers of satisfaction – reliability, transparency, and perceived control – thus appear to operate similarly across organizational contexts.

These results carry several implications. For platform providers developing tools for AI and ML development, they underscore the importance of designing MLOps features that scale across different organizational contexts. Capabilities such as workflow orchestration, model monitoring, and experiment tracking should be adaptable to varying levels of complexity – lightweight enough for smaller teams yet comprehensive enough for enterprise deployment. For organizations, the results suggest that smaller firms may benefit from adopting MLOps practices earlier in their growth trajectory to manage increasing coordination demands, while larger firms should focus on integration and governance across distributed teams. In both cases, satisfaction depends less on firm size itself and more on how effectively MLOps practices are embedded into day-to-day development processes.

Together, these findings refine our understanding of the role of organizational context in information systems success. Firm size shapes how much users talk about MLOps practices but not how much they value them. MLOps thus emerges as a cross-contextual success factor – its benefits for coordination, reliability, and transparency are not confined to large enterprises but extend across the full spectrum of organizational scales.

### 5.3. Limitations and Future Research

This study provides one of the first large-scale empirical assessments of how MLOps best practices relate to user satisfaction with machine learning development platforms. While the findings offer new insight into how operational and organizational dimensions of MLOps shape user evaluations, several limitations should be acknowledged that also point to promising directions for future research.

A first limitation lies in the reliance on product reviews as the primary data source. This approach enables scalable, real-world observation of how practitioners describe their experiences with AI development platforms, but it also introduces potential sampling and reporting biases. Reviewers who post on G2 may differ systematically from the broader user population – for example, in expertise, motivation, or organizational role. Moreover, the G2 user base is skewed toward commercially oriented platforms, which may underrepresent open-source or internal enterprise environments. Future studies could address these limitations by combining review data with survey or usage-based data, or by conducting cross-platform comparisons to assess how perceptions of MLOps differ across contexts and user groups.

A second limitation concerns the use of zero-shot classification with the Llama 3.3 model to infer sentiment toward nine established MLOps practices. Although validation against human annotations showed substantial agreement, LLM-based classification inevitably involves probabilistic judgment and may emphasize linguistic salience rather than underlying intent or implementation depth. Our operationalization identifies whether practices are mentioned and the direction of sentiment, but not the extent, maturity, or practical enactment of those practices. Future work could improve construct measurement through supervised or few-shot learning approaches, expert-coded labels, or multi-level scales that capture both the presence and the perceived maturity of MLOps capabilities.

A further limitation relates to the exclusive use of textual data to assess MLOps practices. While user-generated texts provide valuable, unprompted insights, they reflect perceptions only when explicitly articulated and may overlook implicit experiences. Textual data also limit the ability to distinguish between individual and organizational perspectives. Future research could integrate textual evidence with structured survey measures, behavioral usage data, or interviews to triangulate how users experience and evaluate MLOps practices.

Moreover, our analysis captures only one dimension of organizational context. Firm size serves as a practical and widely used proxy for organizational scale and complexity, and it was readily available in the G2 metadata. However, it does not reflect other contextual characteristics – such as industry, digital maturity, governance model, or structural configuration – that may also influence how MLOps practices are implemented and perceived. Future research could incorporate these additional dimensions to develop a more nuanced understanding of how organizational environments shape the adoption and impact of MLOps.

Finally, this study analyzes MLOps from the user perspective. Complementary firm-level analyses could examine how the adoption and institutionalization of MLOps practices affect broader organizational outcomes such as innovation capability, deployment frequency, data governance, or operational efficiency. Linking user-level perceptions with firm-level practices would allow future work to more fully capture how MLOps maturity translates into both technical and organizational performance.

## 6. Conclusion

This study provides new empirical evidence on how MLOps best practices shape user satisfaction with machine learning development platforms. Drawing on large-scale user reviews and zero-shot text classification, we show that positive perceptions of MLOps practices are systematically associated with higher satisfaction, underscoring that these practices are not merely technical ideals but integral components of users' experience with AI development tools. The results further reveal that reviewers from small firms discuss several MLOps practices less often than those from large enterprises, suggesting differences in exposure and organizational maturity, while the benefits of MLOps – once experienced – appear largely universal across firm sizes.

Together, these findings highlight MLOps as a meaningful lever for both platform providers and adopting organizations. For platform providers, they emphasize the importance of embedding and communicating robust MLOps capabilities as part of the user experience. For organizations developing and deploying AI, they suggest that investing in MLOps practices can enhance not only operational reliability and transparency but also user trust and satisfaction. By linking MLOps to established theories of information processing, technology fit, and system success, this study contributes to a growing understanding of how the operational foundations of AI development translate into perceived value and success in practice.

# Appendix

## Appendix A: Prompt for LLM classifications

> **Classify the following user review of an AI product based on the following nine MLOps practices:**
>
> **1. CI/CD automation: [Definition from Kreuzberger et al., 2023]**
> **2. Workflow orchestration: [Definition from Kreuzberger et al., 2023]**
> **3. Reproducibility: [Definition from Kreuzberger et al., 2023]**
> **4. Versioning: [Definition from Kreuzberger et al., 2023]**
> **5. Collaboration & Communication: [Definition from Kreuzberger et al., 2023]**
> **6. Continuous ML training & evaluation: [Definition from Kreuzberger et al., 2023]**
> **7. ML metadata tracking/logging: [Definition from Kreuzberger et al., 2023]**
> **8. Continuous monitoring: [Definition from Kreuzberger et al., 2023]**
> **9. Feedback loops: [Definition from Kreuzberger et al., 2023]**
>
> **For each practice, assign a value based on the content of the review:**
> - **0: The practice is not discussed in the review.**
> - **1: The practice is discussed in a positive way.**
> - **-1: The practice is discussed in a negative way.**
> - **2: The practice is mentioned in a neutral way.**
>
> **Please analyze the sentiment of the review for each of these practices and provide your output as a list of numbers in the following format: [CI/CD automation, Workflow orchestration, Reproducibility, Versioning, Collaboration & Communication, Continuous ML training & evaluation, ML metadata tracking/logging, Continuous monitoring, Feedback loops]**
>
> **Example Output:**
> **For a given user review, the output might look like this: [1, 0, -1, 0, 0, 1, 0, 2, -1]**
> **(This would indicate that Automating CI/CD, Versioning, and Continuous training were discussed positively; Reproducibility and Feedback loops were discussed negatively; Continuous monitoring was mentioned in a neutral way; and the rest were not discussed.)**
>
> **Only respond in the following form: [X, X, X, X, X, X, X, X, X]. Do not provide any extra explanation or deviate from the targeted output format.**

Note: The definitions for each MLOps practice were adapted verbatim from Kreuzberger et al. (2023). To comply with copyright restrictions, they are not reproduced here but were included in the original prompt when used for classification.

# Appendix B: Examples of MLOps Sentiment Classifications

## Table AT1: Examples of MLOps Classifications

| MLOps Practice | Positive Sentiment | Negative Sentiment |
|---|---|---|
| CI/CD | Like: [...] Deployments are now breeze, From Data Scientist perspective its switch and play. Very low involvement of Devops Team in whole process.<br>2. Platform provides an easy to Use UI. Very easy to deploy services with security inbuilt.<br>3. Cloud Agnoistic in nature.<br>4. Reduced deployment/maintanence time by 80%. | Dislike: [...] From a deployment perspective, the application is truly a nightmare. It's extremely archaic in the sense that it expects long-standing configuration files to be stored somewhere locally and that it requires an internet connection. [...] |
| Workflows | Like: I can see all my pipeline in one place and know exactly what I need to do to hit my number. I build all my own content as well. | Dislike: does not have a built-in scheduler which would be very helpful for workflow orchestration and automation |
| Reproducibility | Like: the platform's data traceability functionality is outstanding, allowing you to track and audit all changes to your data and models, ensuring transparency and reproducibility. | Dislike: The engine is unpredictable resulting in different results when using it for the same workflow |
| Versioning | Like: Model versioning for MLOps. its a great tool for model life cycle management | Dislike: Version compare feature could be made more robust. [...] |
| Collaboration | Like: Most helpful Is the collarobation features. It is simply the only current company that provides seemless collaboration with multiple users. | Dislike: [...] We have stakeholders in both technical and business areas, and struggle to communicate this across teams. [...] |
| Cont. Training | Like: [...] One notable feature is the base model tuning, allowing users to tailor models according to specific requirements. This flexibility ensures that the AI models generated are not only powerful but also customized to meet the unique demands of different projects. | Dislike: Scenario forecasting is something we urge to have. More white-labelling features. Model to reatain training and not to reboot at each conversation session. The possibility to have a central learning system (gathering learning from all conversations we have). |
| Metadata | Like: Built-in support for ML metadata, model versioning, model monitoring, explainability, pipelines, and more allows you to industrialize and scale your ML projects. Less DevOps work for your team. [...] | Dislike: intransparent display of results - which result belongs to which run - how can i compare my old result - bad naming for them as well |
| Monitoring | Like: The ease to deploy an AI model, with out-of-the-box explainability and the necessary governance & compliance tools and monitoring functionalities. […] | Dislike: it is slow sometimes while making annotations, lack of realtime notifications |

**Table AT1. Continued**

| Fedback Loops | Like: The option of serving our models on their platform gives us the flexibility to use it as a 'model proof-of-concept / validation system' with a human in the loop that helps to uncover blindspots and edge cases in our production models. […] | Dislike: would be great to have more options for triggering automation based on specific data events. It's understandable |
|---|---|---|